\documentstyle[12pt]{article}
\newlength{\mathspace}
\def\np#1{ Nucl. Phys. B#1}
\def\pr#1    { Phys. Rev. D#1 }
\def\pl#1{ Phys. Lett. B#1}
\def\cmp   { Commun. Math. Phys. }

\def\ijmp#1  { Int. Jour. Mod. Phys. A#1 }
\def\mpl#1   { Mod. Phys. Lett. A#1 }
\def\begineq{\begin{equation}}
\def\endeq{\end{equation}}
\def\eqabegin{\begin{eqnarray}}
\def\eqaend{\end{eqnarray}}
\def\nn{\nonumber}

\def\parbigskip        {  \par\bigskip  }
\def\parmedskip        {  \par\medskip  }

\def\parbigskipn        {  \par\bigskip\noindent  }

\begin{document}
\baselineskip=0.7cm
\setlength{\mathspace}{2.5mm}



\begin{titlepage}

    \begin{normalsize}
     \begin{flushright}
                 MRI-PHY/14/95\\
                 US-FT-19/95 \\
                 hep-th/9507054 \\
     \end{flushright}
    \end{normalsize}
    \begin{LARGE}
       \vspace{1cm}
       \begin{center}
         {BRST Cohomology Ring in ${\hat c_M}<1$}\\
         {NSR String Theory} \\
       \end{center}
    \end{LARGE}

  \vspace{5mm}

\begin{center}
           Sudhakar P{\sc anda}
           \footnote{E-mail address:
              panda@mri.ernet.in}

           \vspace{2mm}

             {\it Mehta Research Institute of Mathematics}\\
             {\it and Mathematical Physics}\\
             {\it 10 Kasturba Gandhi Marg, Allahabad 211 002, India}\\

           \vspace{.5cm}

                \ \  and \ \

           \vspace{.5cm}

           Shibaji R{\sc oy}
           \footnote{E-mail address:
              roy@gaes.usc.es}

                 \vspace{2mm}

        {\it Departamento de F\'\i sica de Part\'\i culas} \\
        {\it Universidade de Santiago}\\
        {\it E-15706 Santiago de Compostela, Spain}\\
      \vspace{1cm}

    \begin{large} ABSTRACT \end{large}
        \par
\end{center}
 \begin{normalsize}
\ \ \ \
The full cohomology ring of the Lian-Zuckerman type operators (states)
in ${\hat c_M}<1$ Neveu-Schwarz-Ramond
(NSR) string theory is argued to be generated by three elements $x$, $y$ and
$w$ in analogy with the corresponding results in the bosonic case. The ground
ring generators $x$ and $y$ are non-invertible and belong to the Ramond sector
whereas the higher ghost number operators are generated by an invertible
element $w$ with ghost number one less than that of the ground ring generators
and belongs to either Neveu-Schwarz (NS) or Ramond (R) sector
depending on whether we consider (even, even) or (odd, odd) series coupled to
$2d$ supergravity. We explicitly construct these operators (states) and
illustrate our result with an example of pure Liouville supergravity.
 \end{normalsize}

\end{titlepage}
\vfil\eject
It is well-known that the structure of the cohomolgy class or the physical
state spectrum of $c_M \leq 1$ string theory becomes significantly different
because of the presence of degenerate Virasoro representation in the matter
sector. Although these special states also known as discrete states have been
first observed in the context of $c_M = 1$ matrix model [1,2],
their true origin has
been discovered in the mathematical analysis of ref.[3] in the continuum
Liouville approach. For $c_M = 1$ string theory these states have been
completely classified and the physical consequences of the existence of such
states have been elaborated in ref.[4--7].
In particular, using the state-operator
correspondence the short distance limit of operator products of certain spin
zero ghost number zero operators were found to generate a commutative,
associative ring. This so-called ``ground ring" structure [6] was useful in
explaining, among other things, the free fermion description of $c_M = 1$
matrix model. In $c_M<1$ string theory, however, it was realized in ref.[8,9]
that one needs an additional invertible element $w$ with ghost number $-1$ to
describe the complete physical state spectrum. So, unlike in $c_M = 1$ case,
there are physical states at an arbitrary ghost number and the ghost number
$`n$'
sector of the physical operators is given as ${\cal R} = {\cal R}_0\,w^{-n}$,
where ${\cal R}_0$ is the ring of ghost number zero operators. Consequently,
the short distance limit of the operator products of such operators defines a
ring structure in the relative cohomology $H_{rel} = \oplus_{n\in Z}
H_{rel}^{(n)}$

In order to have a better understanding of a more realistic string theory, we
need to know the cohomology structure of ${\hat c}_M \leq 1$ NSR string theory.
This program has been taken in ref.[10--12], but the structure of the complete
physical state spectrum for ${\hat c}_M<1$ NSR string has not been reported
in the literature. In this paper, in analogy with the bosonic case, we first
argue kinematically that the full spectrum of ${\hat c}_M < 1$ NSR string
theory can be generated by two ground ring generators $x$ and $y$ with one
more additional generator $w$. The ground ring generators are non-invertible
and ghost number\footnote[2]{There are some subtleties concerning the ghost
number of a physical operator which are explained later.} $-\frac{1}{2}$
operators which belong to the R-sector. We explicitly
construct these operators using Lian-Zuckerman theorem for the supersymmetric
case [3,11].
We then note that the ground ring contains a subring generated by three
elements in NS-sector with an equivalence relation among them. On the other
hand the generator $w$ has ghost number one less than that of the ground ring
generators and is invertible. Although it
is easy to write $w^{-1}$ for a general $(p,q)$ model coupled to $2d$
supergravity, the construction of $w$ becomes quite involved since one has to
deal with higher level null vectors. But, in general we can say that $w$ will
appear either in NS-sector or in R-sector depending on whether we choose $p$,
$q$ to be even or odd. We, therefore, choose pure Liouville supergravity
or (2, 4) model coupled to $2d$ supergravity as an illustrative example and
construct $w$ explicitly for this model. In this case $w$ belongs to
NS-sector. By using picture changing isomorphism we then show that $w$ is
indeed invertible. Thus we obtain that the physical operators at ghost number
`$n$' sector upto a constant shift, is given by
${\cal R} = {\cal R}_0\,w^{-n}$ and therefore, the
complete cohomology ring structure of ${\hat c}_M < 1$ NSR string theory
coincides with that of the bosonic case.

The supersymmetric minimal models [13--17] coupled to $2d$ supergravity can be
described in terms of Coulomb gas representation with the energy-momentum
tensor and the supercurrent having the form \footnote [1]{Since we are working
in the free theory with zero cosmological constant we will concentrate only
on the holomorphic sector. Free boson and fermion propagators are taken as
$\langle \phi(z) \phi(w) \rangle = - \log(z-w)$ and $\langle \psi(z) \psi(w)
\rangle = \frac{1}{z-w}$.},
\eqabegin
T &=& T^M + T^L\nn\\
     &=& -\frac{1}{2}:\partial\phi_M\partial\phi_M: + i{\hat Q}_M
         \partial^2\phi_M - \frac{1}{2}:\psi_M\partial\psi_M: +
         (M\leftrightarrow L)\\
G &=& G^M + G^L\nn\\
     &=& \frac{i}{2}\partial\phi_M\psi_M + {\hat Q}_M\partial\psi_M
         + (M\leftrightarrow L)
\eqaend
Here $\big(\psi_M(z), \phi_M(z)\big)$ and $\big(\psi_L(z), \phi_L(z)\big)$ are
the $N=1$ matter and Liouville supermultiplets. $2{\hat Q}_M$ and
$2{\hat Q}_L$ are the background charges for matter and Liouville sector. The
matter and Liouville central charges are $\frac{2}{3} c_M \equiv {\hat c}_M
= 1 - 8 {\hat Q}_M^2$ and  $\frac{2}{3} c_L \equiv {\hat c}_L
= 1 - 8 {\hat Q}_L^2$. Since the total central charge of the combined
matter-Liouville system $c_M + c_L$ is equal to 15, the background charges
satisfy
\begineq
{\hat Q}_M^2 + {\hat Q}_L^2 = -1
\endeq
The matter sector in this case is ${\hat c}_M < 1$ superminimal models
which are characterized by Virasoro central charge
\begineq
{\hat c}_M  = 1 - \frac{2(p-q)^2}{pq}
\endeq
where $p,\, q \in 2Z+1$ and gcd$(p,\,q) = 1$ or $p,\, q \in 2Z$ with gcd$(
\frac{p}{2},\,\frac{q}{2}) = 1$ and $\frac{1}{2}(p-q)$ = odd, and so,
the background
charges of the matter and Liouville sector can be parametrized using (4) and
(3) as,
\eqabegin
{\hat Q}_M &=& \frac{1}{2}\left(\sqrt {\frac{p}{q}} - \sqrt {\frac{q}{p}}
\right)\\
{\hat Q}_L &=& \frac{i}{2}\left(\sqrt {\frac{p}{q}} + \sqrt {\frac{q}{p}}
\right)
\eqaend
The vertex operators $e^{i p_M \phi_M}$ and $e^{i p_L \phi_L}$ have conformal
weights $\Delta(p_M) = \frac{1}{2} p_M(p_M - 2 {\hat Q}_M)$ and
$\Delta(p_L) = \frac{1}{2} p_L(p_L - 2 {\hat Q}_L)$ respectively. Equating
them to $\frac{1}{2}$, we get the screening charges of the matter and
Liouville sector as follows,
\eqabegin
\alpha_{\pm} &=& p_M^{\pm} \,=\, {\hat Q}_M \pm \sqrt {{\hat Q}_M^2 + 1}\\
\beta_{\pm} &=& p_L^{\pm} \,=\, {\hat Q}_L \pm \sqrt {{\hat Q}_L^2 + 1}
\eqaend
We note from (7) that, $\alpha_+ + \alpha_- = 2 {\hat Q}_M \equiv \alpha_0,
\,\alpha_+\,\alpha_- = -1$ and similarly from (8), $\beta_+ + \beta_- =
{\hat Q}_L \equiv \beta_0,\, \beta_+\,\beta_- = -1$. So, comparing with (5)
and (6), we find
\eqabegin
\alpha_+ &=& \sqrt{ \frac{p}{q}} \,=\, -i\beta_+\\
\alpha_- &=& -\sqrt{ \frac{q}{p}} \,=\, i\beta_-
\eqaend
In this notation the matter momenta can be parametrized as $\alpha_{m,m'}
= \frac{1}{2}\left[(1-m)\alpha_+ + (1-m')\alpha_-\right]$ where $1\leq m
\leq q-1$, $1\leq m' \leq p-1$ and the conformal weight of the corresponding
vetex operator is given by,
\begineq
\Delta(\alpha_{m,m'}) = \frac{(pm - qm')^2 - (p-q)^2}{8pq} + \frac{1}{32}
\left[ 1 - (-1)^{m-m'} \right]
\endeq
where $m-m' \in 2Z$ for NS representation and $m-m' \in 2Z+1$ for
R-representation.

The irreducible highest weight module of the matter sector is obtained by
quotienting the Verma module over each primary by its maximum proper
submodule. Like in the bosonic case, in the supersymmetric case also the
submodule is generated by a pair of null vectors at level $\frac{mm'}{2}$
and $\frac{(p-m')(q-m)}{2}$ associated with the primary field. The irreducible
super-Virasoro module can be represented by the following embedding diagram
$E_{m,m'}$:
\begineq
\def\sp{\nearrow\!\!\!\!\!\!\searrow}
\matrix{&&a_{-1}&\longrightarrow&e_{-1}&\longrightarrow&a_{-2}
&\longrightarrow&e_{-2}&\longrightarrow&\cdots\cr
e_0&\nearrow\atop\searrow&&\sp&&\sp&&\sp&&\sp&\cdots\cr
&&a_0&\longrightarrow&e_1&\longrightarrow&a_1&\longrightarrow&e_2
&\longrightarrow&\cdots\cr}
\endeq
where
\eqabegin
a_t &=& \frac{1}{8pq}\left[(2pqt+pm+qm')^2 - (p-q)^2\right] + \frac{1}{32}
\left[1-(-1)^{m-m'}\right]\nn\\
e_t &=& \frac{1}{8pq}\left[(2pqt+pm-qm')^2 - (p-q)^2\right] + \frac{1}{32}
\left[1-(-1)^{m-m'}\right]
\eqaend
Each node of the above diagram represents a Verma module with $a_t$ and $e_t$
as the dimension of its highest weight state and are null vectors over $e_0$.
An arrow connecting two spaces $E \to F$ means that the module $F$ is contained
in module $E$. It is stated by Lian and Zuckerman [3] and also by Bouwknegt,
McCarthy and Pilch [11], under the plausible assumption
that the Felder resolution [18]
generalizes to the supercase, that there exists a unique physical state of the
combined matter, Liouville and ghost system at each of these values of $a_t$
and $e_t$ if and only if the Liouville momenta satisfy the relation
\begineq
\Delta(p_L) = \cases {\frac{1}{2} - a_t \cr \frac{1}{2} - e_t\cr}
\endeq
and the ghost number of the state is given as\footnote[2]{The ghost number of
a state is defined with the convention that the highest weight ghost vacuum
has ghost number zero.},
\begineq
n = \pi(p_L) d(p_L)
\endeq
where $\pi(p_L)$ = sign $[i(p_L-{\hat Q}_L)]$ and $d(p_L)$ is the number
of arrows from the top node $e_0 \equiv \Delta(\alpha_{m,m'})$ to that
particular node $a_t$ or $e_t$.

In the BRST quantization scheme, the states are taken to be in the space of
conformal fields such that a physical state is BRST closed and belong to the
zero eigenvalue space of $L_0^{tot}$. The nilpotent BRST charge is defined as,
\begineq
Q_{BRST} = \oint\,dz :\left[c\left(T^M + T^L + \frac{1}{2} T^{gh}\right)
           + \gamma\left(G^M + G^L + \frac{1}{2} G^{gh}\right)\right]:
\endeq
where $T^M(z)$, $T^L(z)$, $G^M(z)$, $G^L(z)$ are as given in Eqs.(1) and (2)
and the ghost energy-momentum tensor and supercurrent are given as\footnote[1]
{The basic OPE's are taken as $b(z)c(z)\sim \frac{1}{z-w};\,\beta(z)\gamma(w)
\sim \frac{1}{z-w}$},
\eqabegin
T^{gh} &=& -2:b\partial c: - :\partial b c: + \frac{3}{2}
:\beta\partial\gamma: + \frac{1}{2}\partial\beta\gamma:\\
G^{gh} &=& -\frac{3}{2}\partial c \beta - c\partial \beta - \frac{1}{2}b\gamma
\eqaend
Here $b(z)$, $c(z)$ are the diffeomorphism ghosts with conformal weights
2 and $-1$,
whereas $\beta(z)$, $\gamma(z)$ are the superdiffeomorphism ghosts with
conformal
weights $\frac{3}{2}$ and $-\frac{1}{2}$ respectively. In terms of modes
$Q_{BRST}$ can be expanded for NS sector as,
\begineq
Q_{BRST} = c_0 L_0^{tot} + M b_0 + {\tilde Q}_{BRST}
\endeq
and for R-sector,
\begineq
Q_{BRST} = c_0 L_0^{tot} + \gamma_0 G_0^{tot} + M b_0 + N \beta_0 +
{\tilde Q}_{BRST}
\endeq
where
\eqabegin
L_0^{tot} &=& L_0^L + L_0^M + \sum_n\,n:c_{-n} b_n: + \sum_r\,
r:\gamma_{-r} \beta_r:\\
G_0^{tot} &=& G_0^L + G_0^M - \frac{1}{2}\sum_n\,n c_{-n}\beta_n - \frac{1}{2}
\sum_n\,\gamma_{-n} b_n\\
M &=& -\sum_n\,n c_{-n} c_n - \frac{1}{4} \sum_r\,\gamma_{-r} \gamma_r\\
N &=& \sum_n\,\frac{3}{2} n \gamma_{-n} c_n\\
{\tilde Q}_{BRST} &=& \sum_{n\neq 0}\,c_{-n}\left(L_n^L + L_n^M\right) +
\sum_{r \neq 0}\,\gamma_{-r}\left(G_r^L + G_r^M\right)+\frac{1}{2}\sum_{
{n,m,} \atop{n+m\neq 0}}\,(m-n):c_{-n} c_{-m} b_{n+m}:\nn\\
& & \qquad\qquad + \sum_{n,r \neq 0}\, \left(r-\frac{1}{2} n\right):c_{-n}
\gamma_{-r}\beta_{n+r}: - \frac{1}{4}\sum_{n,r \neq 0}\,\gamma_{-r} b_{-n}
\gamma_{n+r}
\eqaend
In Eqs.(21), (23), and (25), $r$ is half-integral for NS-sector and integral
for
R-sector. Also, in terms of oscillators  the super-Virasoro generators are
given below,
\eqabegin
L_n^{L,M} &=& \frac{1}{2}\sum_m\,:\alpha_m^{L,M}\alpha_{n-m}^{L,M}: - (n+1)
{\hat Q}_{L,M}\,\alpha_n^{L,M} - \frac{1}{2}\sum_m\,\left(m+\frac{1}{2}\right)
:\psi_m^{L,M}\psi_{n-m}^{L,M}:\\
G_n^{L,M} &=& \frac{1}{2}\sum_m\,\alpha_m^{L,M}\psi_{n-m}^{L,M}
-\left(n+\frac{1}{2}\right){\hat Q}_{L,M}\,\psi_n^{L,M}
\eqaend
where $L,\,M$ refer to Liouville and matter sectors respectively.
In the space of conformal fields a general physical operator takes the form
\begineq
{\cal O} = {\cal P}\left(\partial\phi_M,\,\partial\phi_L,\,\psi_M\,\psi_L\,
b,\,c,\,\beta,\,\gamma\right)\,e^{i\alpha_{m,m'}\phi_M +i\beta_{n,n'}\phi_L}
\endeq
where, because of the zero eigenvalue of $L_0^{tot}$, ${\cal P}$ is a
differential polynomial of conformal weight $-\frac{1}{2}\alpha_{m,m'}\left(
\alpha_{m,m'} - 2 {\hat Q}_M\right)-\frac{1}{2}\beta_{n,n'}
\left(\beta_{n,n'} - 2 {\hat Q}_L\right)$. The Liouville momentum
$\beta_{n,n'}$ is defined as usual by $\beta_{n,n'} = \frac{1}{2}\left[(1-n)
\beta_+ + (1-n')\beta_-\right]$ with $\beta_{\pm}$ as given in (8).
We also note
the following symmetry properties of the momenta as $\alpha_{m,m'} = \alpha_
{m\pm q, m'\pm p}$ and $\beta_{n,n'} = \beta_{n\pm q, n'\mp p}$. Finally,
adding momenta we get, $\alpha_{m,m'} + \alpha_{n,n'} = \alpha_{m+n-1,
m'+n'-1}$ and $\beta_{m,m'} + \beta_{n,n'} = \beta_{m+n-1,m'+n'-1}$

With this preparation, we first like to comment that the ring structure for the
operators in the BRST cohomology is quite obvious. Since a physical operator
would have to belong to the zero eigenvalue space of $L_0^{tot}$, it is clear
that the short distance limit of two physical operators can give non-trivial
result only in the regular term. Including the higher ghost number operators
we, therefore, get a commutative and associative ring structure,
\begineq
H^{(n)}_{rel} \times H^{(m)}_{rel} \rightarrow H^{(n+m)}_{rel}
\endeq
In the above `$rel$' implies that we concentrate only on the relative
cohomology
i.e. the physical operators would belong to the zero eigenvalue space of
$L_0^{tot}$, $b_0$ for NS-sector and to the zero eigenvalue space of
$G_0^{tot}$, $b_0$, $\beta_0$ for R-sector. Since the OPE of the vertex
operators of a free scalar field is given by
\begineq
:e^{i\alpha \phi(z)}:\,:e^{i\beta \phi(z)}:\, \sim (z-w)^{\alpha\beta}\,
:e^{i\alpha \phi(z) + i\beta \phi(w)}:
\endeq
we get individual momentum conservation of the matter and Liouville sector
under the ring multiplication. By making use of this fact and by looking at the
possible values of the Liouville momenta of physical operators (28) dictated by
the Lian-Zuckerman theorem (14) we argue that all the physical operators of
${\hat c}_M < 1$ NSR string theory can be generated by three elements in the
cohomology. By solving Eq.(14), we indeed note that the allowed Liouville
momenta can be split into three parts, each part associated with one generator,
which can be summarized as follows,
\eqabegin
p^+_L(a_{-t}) &=& \beta_{-qt+m,-pt+m'}\,=\,-2t\beta_w + (m'-1)\beta_x + (m-1)
\beta_y\\
p^+_L(a_{t-1}) &=& \beta_{-q(t-1)-m,-p(t-1)-m'}\,=\,-2t\beta_w +
(p-m'-1)\beta_x + (q-m-1)\beta_y\\
p^+_L(e_{-t}) &=& \beta_{-qt+m,-pt-m'}\,=\,-(2t+1)\beta_w + (p-m'-1)\beta_x
+ (m-1)\beta_y\\
p^+_L(e_{t}) &=& \beta_{-qt-m,-pt+m'}\,=\,-(2t+1)\beta_w + (m'-1)\beta_x +
(q-m-1)\beta_y\\
p^+_L(e_0) &=& \beta_{-m,m'}\,=\,-\beta_w + (m'-1)\beta_x + (q-m-1)\beta_y\\
p^-_L(a_{-t}) &=& \beta_{qt-m,pt-m'}\,=\,(2t-2)\beta_w + (p-m'-1)\beta_x +
(q-m-1)\beta_y\\
p^-_L(a_{t-1}) &=& \beta_{q(t-1)+m,p(t-1)+m'}\,=\,(2t-2)\beta_w +
(m'-1)\beta_x + (m-1)\beta_y\\
p^-_L(e_{-t}) &=& \beta_{qt-m,pt+m'}\,=\,(2t-1)\beta_w + (m'-1)\beta_x +
(q-m-1)\beta_y\\
p^-_L(e_t) &=& \beta_{qt+m,pt-m'}\,=\,(2t-1)\beta_w + (p-m'-1)\beta_x +
(m-1)\beta_y\\
p^-_L(e_0) &=& \beta_{m,-m'}\,=\,-\beta_w + (p-m'-1)\beta_x + (m-1)\beta_y
\eqaend
where $t\in Z_+$ and we have defined
\eqabegin
\beta_x &=& \beta_{1,2}\,=\,-\frac{i}{2}\sqrt{\frac{q}{p}}\nn\\
\beta_y &=& \beta_{2,1}\,=\,-\frac{i}{2}\sqrt{\frac{p}{q}}\\
\beta_w &=& p \beta_x\,=\,q\beta_y\,=\,\beta_{1,p+1}\,=\,\beta_{q+1,1}\,=\,
-\frac{i}{2}\sqrt{pq}\nn
\eqaend
We would like to comment that the coefficients of $\beta_w$ terms in
(31--40) is
precisely the negative of [the ghost number of the operator $+\frac{1}{4}
(3+(-1)^{m-m'})$] as calculated from
the Lian-Zuckerman theorem (15). This additional shift in the ghost number
by $\frac{1}{4}(3+(-1)^{m-m'})$ appears when we convert a state to an operator.
To be precise, the ghost number of an operator is equal to the ghost number of
the state plus $\frac{1}{4}(1-(-1)^{m-m'})$. This will become clear later
when we describe our convention for the ghost vacuum for NS and R-sector.
It should be emphasized here that the observation that the coefficients of
$\beta_w$ term is related to the ghost number of the operator upto a constant
shift is valid only if we use the picture changing isomorphism to be discussed
later. It is also quite clear that a general operator of ghost number
$\Big[n+1+\frac{3}{4}(-)^i+\frac{3}{4}(-)^j+(-)^{i+j+1}
+\frac{3}{2}(-)^{ij+1}\Big]$
in the cohomology can be written upto BRST-exact terms as,
\begineq
{\cal O} = w^{-n}\,x^i\,y^j
\endeq
where $\Big[\frac{3}{4}(-)^i+\frac{3}{4}(-)^j+(-)^{i+j+1}
+\frac{3}{2}(-)^{ij+1}\Big]$
is the ghost number of the operator $x^iy^j$ upto picture changing isomorphism
(see later for explanation).
Also $i$, $j$ are restricted as $0\leq i\leq p-2$, $0\leq j\leq q-2$ because
we want the matter momenta to lie inside the Kac-table and $n \in Z$.
{}From (37) we note that the elements $x$ and $y$ will appear at $a_0$
in the embedding diagram with matter momenta $\alpha_{1,2}$ and $\alpha_{2,1}$
respectively. Also, the ghost number of $x$ and $y$ is $-\frac{1}{2}$.
It is clear from (35) and (40) that $w^{-1}$
will appear at $e_0$ in the embedding diagram with matter and Liouville momenta
either $(\alpha_{q-1,1},\,\, \beta_{-q+1,1})$ for $p_L > {\hat Q}_L$ or
$(\alpha_{1,p-1},\,\, \beta_{1,-p+1})$ for $p_L < {\hat Q}_L$. Finally, from
(38) and (39) we note that $w$ will appear either at $e_{-1}$ with matter and
Liouville momenta $(\alpha_{q-1,1},\,\, \beta_{1,p+1})$ or at $e_1$ with
momenta
$(\alpha_{1,p-1},\,\, \beta_{q+1,1})$ for $p_L < {\hat Q}_L$. From the above
observation, it is clear that since $m-m'$ is odd for $x$ and $y$, they will
belong to R-sector, whereas for $w$, $w^{-1}$, $(m-m')$ is either $p$ or $q$.
So, for (odd, odd) series they will belong to R-sector and for (even, even)
series they will belong to NS-sector.

Before we describe the construction of the ground ring generators, we give
below our convention for the vacuum in NS and R-sector. The highest weight
ghost vacuum state [19] for NS-sector has the form
\begineq
:c(0)\,e^{\phi(0)}:|0\rangle
\endeq
and the highest weight ghost vacuum for R-sector is given as,
\begineq
:c(0)\,e^{\phi(0)/2}:|0\rangle
\endeq
where $|0\rangle$ is the super-SL$_2$ invariant vacuum and $\phi(z)$ is a
bosonic field obtained by bosonizing the superdiffeomorphism ghost system
as\footnote[1]{Note that our convention of bosonization is slightly different
than the convention used in ref.[19].},
\eqabegin
\beta(z) &=& :\eta(z)\,e^{\phi(z)}:\nn\\
\gamma(z) &=& :\partial\xi(z)\,e^{-\phi(z)}:
\eqaend
In our convention ghost number of the states (43) and (44) are zero. In this
convention, it is clear that there will be a shift of ghost number of $\frac
{1}{4}(1-(-)^{m-m'})$ when we pass from state to operator.
The conformal weight of (43) is
$-\frac{1}{2}$ as expected, but the conformal weight of (44) is $-\frac{5}{8}$.
An additional $\frac{1}{8}$ in R-sector comes from the fact that the matter
and the Liouville sector contains spin fields of weights $\frac{1}{16}$
each [19].
The vacuum for the matter-Liouville sector in this case is degenerate and we
denote them as
\begineq
|\pm\rangle = S^{\pm}(0)|0\rangle
\endeq
where $S^{\pm}(z)$ are the spin fields with conformal weight $\frac{1}{8}$. In
our convention \newline $(\psi_0^M + i\psi_0^L)|+\rangle = 0$ and
$(\psi_0^M - i\psi_0^L)|+\rangle = |-\rangle$. Finally, in R-sector an
additional complication arises because of the fact that the relative cohomology
states should be annihilated by $G_0^{tot}$. This can be done either by solving
explicitly the following equations,
\eqabegin
G_0^{tot}|{\rm state}\rangle &=& 0\nn\\
{\tilde Q}_{BRST}|{\rm state}\rangle &=& 0
\eqaend
or by first writing down the state in terms of rotated oscillators [20,11]
and then solving
\begineq
{\tilde Q}_{BRST}|{\rm state(rotated \,\, oscillators)}\rangle = 0
\endeq
where ${\tilde Q}_{BRST}$ is as given in (25) and a rotated oscillator is
defined as
\begineq
{\tilde O}_m = [G_0^{tot}, \Theta\,O_m\} = O_m - \Theta [G_0^{tot}, O_m\}
\endeq
where $\Theta = \frac{1}{2}(\psi_0^M + i\psi_0^L)/[(p_M - {\hat Q}_M) +
i (p_L - {\hat Q}_L)]$ and $\{G_0^{tot}, \Theta\} = 1$. It is easy to verify
that the rotated oscillators satisfy the same algebra as the original ones
and the states built out of rotated oscillators are automatically annihilated
by $G_0^{tot}$. Since the ground ring generators contain oscillators at level
one, by writing down states in terms of rotated oscillators and then solving
(48) we obtain,
\eqabegin
x &=& \Big[b_{-1} + 2 \beta_{-1}\left(G_0^L + G_0^M \right)\Big]|h_{1,2}^M;\,
h_{1,2}^L;\, +\rangle\\
y &=& \Big[b_{-1} + 2 \beta_{-1}\left(G_0^L + G_0^M \right)\Big]
\left(\psi_0^M - i\psi_0^L\right) |h_{2,1}^M;\, h_{2,1}^L;\, +\rangle
\eqaend
where
\eqabegin
|h_{1,2}^M;\, h_{1,2}^L;\, +\rangle &=& :e^{i \alpha_{1,2} \phi_M(0)
+ i \beta_{1,2}
\phi_L(0)}\,c(0)\,e^{\phi(0)/2}\,S^+(0):|0\rangle\nn\\
|h_{2,1}^M;\, h_{2,1}^L;\, +\rangle &=& :e^{i \alpha_{2,1} \phi_M(0)
+ i \beta_{2,1}
\phi_L(0)}\,c(0)\,e^{\phi(0)/2}\,S^+(0):|0\rangle
\eqaend
In obtaining (50) and (51), we encounter the null vectors at level one for both
matter and Liouville sectors. They are given as,
\eqabegin
& &\left(L_{-1}^M - \frac{8}{(2 - p/q)} G_{-1}^M G_0^M \right)|h_{1,2}^M;\,
+\rangle\\
& &\left(L_{-1}^M - \frac{8}{(2 - q/p)} G_{-1}^M G_0^M \right)|h_{2,1}^M;\,
+\rangle\\
& &\left(L_{-1}^L - \frac{8}{(2 + p/q)} G_{-1}^L G_0^L \right)|h_{1,2}^L;\,
+\rangle\\
& &\left(L_{-1}^L - \frac{8}{(2 + q/p)} G_{-1}^L G_0^L \right)|h_{2,1}^L;\,
+\rangle
\eqaend
The ground ring generators written as operators have the following form,
\eqabegin
x &=& \left[S^+\,e^{\phi/2} + \sqrt{\frac{q}{p}} S^-\,\eta\,c\,e^{3\phi/2}
\right]e^{i\alpha_{1,2}\phi_M + i\beta_{1,2}\phi_L}\\
y &=& \left[S^-\,e^{\phi/2} + 2\sqrt{\frac{p}{q}} S^+\,\eta\,c\,e^{3\phi/2}
\right]e^{i\alpha_{2,1}\phi_M + i\beta_{2,1}\phi_L}
\eqaend
{}From the explicit form of the ground ring generators (57) and (58) it is
clear
that they have ghost numbers $-\frac{1}{2}$.
The reason that $x$ and $y$ are non-invertible can be understood easily by
noticing that there are no operators in the cohomology whose momenta are
negative of those of $x$ and $y$. Also in order to show that the powers of
$x$ and $y$ give non-trivial cohomology one can proceed exactly in the similar
way as described in ref.[6,7,11] through construction of currents which act as
derivations on $x$ and $y$. Finally, we note that there is a subring of this
ground ring generated by three elements $x^2$, $y^2$ and $xy$ in NS-sector
with a restriction that only odd powers of the third element is allowed, since
the even powers can be identified with some powers of the product of first
two elements. The explicit forms of these operators are given below:
\eqabegin
x_{NS}
& &\,=\,\left[\partial\eta c e^{2\phi} + \frac{1}{2}\sqrt{
\frac{p}{q}}\left(\partial\phi_L - i\partial\phi_M\right) \eta c e^{2\phi}
\right.\nn\\
& &\qquad\qquad\qquad\qquad\left. -\frac{i}{4}\sqrt{\frac{p}{q}}\left(\psi_L
- i\psi_M\right) e^{\phi}\right] e^{i\alpha_{1,3}\phi_M + i\beta_{1,3}\phi_L}\\
y_{NS}
& &\,=\,\left[\partial\eta c e^{2\phi} + \frac{1}{2}\sqrt{
\frac{q}{p}}\left(\partial\phi_L + i\partial\phi_M\right) \eta c e^{2\phi}
\right.\nn\\
& &\qquad\qquad\qquad\qquad\left.-\frac{i}{4}\sqrt{\frac{q}{p}}\left(\psi_L
+ i\psi_M\right) e^{\phi}\right] e^{i\alpha_{3,1}\phi_M + i\beta_{3,1}\phi_L}\\
z_{NS}
& &\,=\,\left[bc e^{\phi} + \frac{p\sqrt{pq}}{p^2+q^2}
\left(\partial\phi_L+i\partial\phi_M \right) e^{\phi} +
\frac{q\sqrt{pq}}{p^2+q^2}\left(\partial\phi_L-i\partial\phi_M \right) e^{\phi}
\right.\nn\\
& &\qquad\qquad -i\frac{p\sqrt{pq}}{p^2+q^2}\left(\partial\psi_L + i\partial
\psi_M\right)\eta c e^{2\phi} - i\frac{q\sqrt{pq}}{p^2+q^2}\left(\partial
\psi_L - i\partial\psi_M\right)\eta c e^{2\phi}\nn\\
& &\qquad\qquad - \frac{2i pq}{p^2+q^2}
\left(\partial\phi_L \psi_L - \partial\phi_M\psi_M\right) \eta c e^{2\phi}
 + i\frac{p\sqrt{pq}}{p^2+q^2}\left(\psi_L + i\psi_M
\right) \partial\eta c e^{2\phi}\nn\\
& &\qquad\qquad\left.+ i\frac{q\sqrt{pq}}{p^2+q^2}
\left(\psi_L - i\psi_M \right)\partial\eta c e^{2\phi}\right]
e^{i\alpha_{2,2}\phi_M + i\beta_{2,2}\phi_L}
\eqaend
The generators $x_{NS}(\equiv x^2)$, $y_{NS}(\equiv y^2)$ and $z_{NS}(\equiv
xy)$ form a subring of the ground
ring modulo the equivalence relation $x_{NS}.y_{NS} \simeq z_{NS}^2$.
We would like to point out that the ghost numbers of the ground ring
generators are $-\frac{1}{4}(3+(-)^{m-m'})$ i.e. $-\frac{1}{2}$ when they
are in R-sector and $-1$ when they are in NS-sector. In fact, we note that
$x_{NS}$, $y_{NS}$ and $z_{NS}$ have ghost numbers $-1$. But in general, under
the ring multiplication they will go outside this set. In order to resolve this
problem, we need to use the picture changing isomorphism [19,21].
The picture changing
operator in our convention has the following form,
\begineq
X = (G^L + G^M) e^{-\phi} + c\eta +\frac{1}{4} b\partial^2\xi e^{-2\phi} +
\frac{1}{4}\partial(b\partial\xi e^{-2\phi})
\endeq
There also exists an inverse of this picture changing operator in the
cohomology denoted as $Y$ whose form is given below,
\begineq
Y = -4 c\eta e^{2\phi}
\endeq
The ghost numbers of $X$ and $Y$ can be checked to be $+1$ and $-1$
respectively. It is also easy to verify that upto BRST-exact term $X^n.Y^n\,
\sim I$. It is well-known that the picture changing operation defines an
isomorphism in the cohomology [21] even in the relative cohomology that we are
interested in. So, using these two operators one can change the ghost number
of an operator to any convenient value. Upto this isomorphism the generators
$x$ and $y$, therefore, define a ring relation under the short distance limit
of the operator products. Also, it is easy to verify that the ghost number of
a general element of the ground ring $x^iy^j$ has the value
$\Big[\frac{3}{4}[(-)^i + (-)^j] + (-)^{i+j+1} + \frac{3}{2}(-)^{ij+1}
\Big]$, upto
picture changing isomorphism, since,
\begineq
\frac{3}{4}[(-)^i + (-)^j] + (-)^{i+j+1} + \frac{3}{2}(-)^{ij+1}\,=\,\cases{
-1 &for {\it i,\,j} = even (NS)\cr
-1 &for {\it i,\,j} = odd (NS)\cr
-\frac{1}{2} &for {\it i} = odd, {\it j} = even (R)\cr
-\frac{1}{2} &for {\it i} = even, {\it j} = odd (R)\cr}
\endeq

To obtain
the rest of the operators we first note that since $w^{-1}$ appears at the top
node $e_0$ of the embedding diagram as discussed earlier it is straightforward
to write
down their form for general $(p,\,q)$ model. For $p$, $q$ even, they are given
as,
\eqabegin
w^{-1} &=& c e^{\phi} e^{i\alpha_{q-1, 1}\phi_M + i\beta_{-q+1, 1}\phi_L}\\
{\rm or} &=& c e^{\phi} e^{i\alpha_{1,p-1}\phi_M + i\beta_{1,-p+1}\phi_L}
\eqaend
and for $p$, $q$ odd, their forms are
\eqabegin
w^{-1} &=& c e^{\phi/2}\,S^+\, e^{i\alpha_{q-1, 1}\phi_M +
i\beta_{-q+1, 1}\phi_L}\\
{\rm or} &=& c e^{\phi/2}\, S^+\, e^{i\alpha_{1,p-1}\phi_M +
i\beta_{1,-p+1}\phi_L}
\eqaend
As pointed out earlier in the discussion after Eq.(42), the ghost number $-2$
operator $w$ will appear either at $e_{-1}$ or at $e_1$ of the embedding
diagram. In general their forms would be when $w$ is in NS-sector,
\eqabegin
w &=&{\cal P}(\partial\phi_M, \partial\phi_L, \psi_M, \psi_L, b, c, \beta,
\gamma) c e^{\phi} e^{i\alpha_{q-1, 1}\phi_M + i\beta_{1, p+1}\phi_L}\\
{\rm or} &=&{\cal P}(\partial\phi_M, \partial\phi_L, \psi_M, \psi_L, b, c,
\beta, \gamma) c e^{\phi} e^{i\alpha_{1,p-1}\phi_M + i\beta_{q+1,1}\phi_L}
\eqaend
or when $w$ is in R-sector
\eqabegin
w &=&{\cal P}(\partial\phi_M, \partial\phi_L, \psi_M, \psi_L, b, c, \beta,
\gamma) c e^{\phi/2} S^+ e^{i\alpha_{q-1, 1}\phi_M + i\beta_{1, p+1}\phi_L}\\
{\rm or} &=&{\cal P}(\partial\phi_M, \partial\phi_L, \psi_M, \psi_L, b, c,
\beta, \gamma) c e^{\phi/2} S^+ e^{i\alpha_{1,p-1}\phi_M +
i\beta_{q+1,1}\phi_L}
\eqaend
where ${\cal P}$ is a differential polynomial of conformal weight $(p+q)/2$.
It can be inferred by observing the momenta of the operators $w$ and $w^{-1}$
in (65--72) that, their product is well-defined if we take $w^{-1}$
from (65) and $w$ from (70) or $w^{-1}$ from (66) and $w$ from (69) when they
are in NS-sector. Also when they are in R-sector their product will be
well-defined when $w^{-1}$ is taken from (67) and $w$ is taken from (72) or
$w^{-1}$ is taken from (68) and $w$ is taken from (71).
We note that for general $(p,\,q)$ models with high values of $p$, $q$
the explicit computation of the operator $w$ gets
quite involved. We, therefore, choose to work in pure Liouville supergravity
for illustration of our results. Since in this case the matter central charge
vanishes, we set the matter fields to zero. Also, we note that in this model
i.e. (2, 4) superminimal model coupled to $2d$ supergravity the operator $w$
belongs to NS-sector. By constructing the most general level three
oscillator with
ghost number $-2$ and then solving ${\tilde Q}_{BRST}|{\rm state}\rangle
= 0$ upto
null vectors, we obtain $|w\rangle$ for this model as,
\eqabegin
|w\rangle &=& \left[b_{-2}b_{-1} - 2 \beta_{-\frac{3}{2}}^2 + \frac{2}{5}
\beta_{-\frac{3}{2}}b_{-1}G_{-\frac{1}{2}}^L - \frac{8}{5} \beta_{-\frac{1}{2}}
b_{-1}G_{-\frac{3}{2}}^L + \frac{4}{5} \beta_{-\frac{1}{2}}b_{-2}
G_{-\frac{1}{2}}^L\right.\nn\\
& &\left. +3 \beta_{-\frac{1}{2}}^3 \gamma_{-\frac{3}{2}} - \frac{8}{5}
\beta_{-\frac{1}{2}} \beta_{-\frac{3}{2}}L_{-1}^L + \frac{1}{2}
\beta_{-\frac{1}{2}}^2 (L_{-1}^L)^2 + \frac{12}{5}
\beta_{-\frac{1}{2}}^2 L_{-2}^L
\right] c e^{\phi} e^{\sqrt {2}\phi_L}|0\rangle
\eqaend
In deriving this result we made use of the null vectors at level 2 and $\frac
{5}{2}$ which are given below,
\eqabegin
& &\left(L_{-2}^L - 2 G_{-\frac{3}{2}}^L G_{-\frac{1}{2}}^L +
\frac{1}{4}(L_{-1}^L)^2\right)|-\frac{5}{2}\rangle\\
& &\left(G_{-\frac{5}{2}}^L + \frac{3}{4}G_{-\frac{3}{2}}^L L_{-1}^L +
\frac{1}{2} G_{-\frac{1}{2}}^L L_{-2}^L + \frac{1}{2} G_{-\frac{1}{2}}^L
(L_{-1}^L)^2\right)|-\frac{5}{2}\rangle
\eqaend
$-\frac{5}{2}$ being the conformal weight of the state $e^{\sqrt{2}\phi_L(0)}
|0\rangle$.
Expressing $w$ as an operator we find,
\eqabegin
w &=& \left[b e^{\phi} + \frac{i\sqrt{2}}{5}\left(2bc\psi_L\eta + \psi_L
\partial\eta + \psi_L\eta\partial\phi + 2\partial\psi_L\eta + 2\sqrt{2}
\partial\phi_L\psi_L\eta\right) e^{2\phi}\right.\nn\\
& & -\left(\frac{1}{3}\partial^3\eta\eta c +\frac{17}{6}\partial^2\eta\partial
\eta c + 5\partial^2\eta\eta c \partial\phi + \frac{13}{2}\partial\eta\eta c
(\partial\phi)^2\right.\nn\\
& & - \frac{9}{2}\partial\eta\eta c \partial^2\phi
 +\frac{4\sqrt{2}}{5}\partial^2\eta\eta c\partial\phi_L
+ \frac{8\sqrt{2}}{5}\partial\eta\eta c
\partial\phi\partial\phi_L +\frac{1}{5}
\partial\eta\eta c (\partial\phi_L)^2\nn\\
& &\left.\left. - \frac{11\sqrt{2}}{10}\partial\eta\eta
c \partial^2\phi_L - \frac{6}{5}\partial\eta\eta c \partial\psi_L\psi_L\right)
e^{3\phi}\right]e^{\sqrt{2}\phi_L}
\eqaend
In order to show that the operator $w$ is invertible i.e. $w.w^{-1} \sim I$
upto BRST-exact terms, we first note that $w^{-1}$ for
(2, 4) model coupled to $2d$ supergravity has the following form,
\begineq
w^{-1} = c e^{\phi} e^{-\sqrt{2}\phi_L}
\endeq
It is quite clear from Eq.(76) that $w^{-1}$ is not in the right ``picture"
since it has ghost number zero. By making use of the picture changing
isomorphism mentioned earlier, we convert $w^{-1}$ to the compatible
picture as given below:

\eqabegin
X.X.w^{-1} &=& \left[-\frac{1}{2\sqrt{2}}\left(\frac{1}{2\sqrt{2}}\partial^2 c
+\partial^2\phi_L c -\partial\phi_L\partial\phi c + \frac{3}{2\sqrt{2}}
\partial^2\phi c - \frac{3}{2\sqrt{2}}(\partial\phi)^2 c\right)e^{-\phi}
\right.\nn\\
& & -\frac{i}{2\sqrt{2}}\left(\frac{1}{2\sqrt{2}}\partial^2\phi_L\psi_L
\partial\xi + \frac{1}{2\sqrt{2}}\partial\phi_L\partial\psi_L\partial\xi
+\frac{3}{4}\partial^2\psi_L\partial\xi +\frac{3}{4}\psi_L\partial^3\xi
\right.\nn\\
& &\qquad\qquad\qquad \left. -3\psi_L\partial\phi\partial^2\xi
- \frac{1}{2}
\psi_L\partial^2\phi\partial\xi + \psi_L(\partial\phi)^2\partial\xi\right)
e^{-2\phi}\nn\\
& & -\frac{1}{4}\left(\frac{1}{4}\partial^2 b\partial^2\xi\partial\xi +
\frac{3}{8}\partial b\partial^3\xi\partial\xi +
\frac{1}{6} b\partial^4\xi\partial\xi - \partial b\partial^2\xi\partial\xi
\partial\phi\right.\nn\\
& &\left.\left. - \frac{3}{4} b\partial^3\xi\partial\xi\partial\phi -
\frac{1}{2}
b\partial^2\xi\partial\xi\partial^2\phi + b\partial^2\xi\partial\xi
(\partial\phi)^2\right)e^{-3\phi}\right] e^{-\sqrt{2}\phi_L}
\eqaend
It is now straightforward to check that $w.(X.X.w^{-1})\,\sim \frac{747}{5}I$.
Thus, by redefining $w$ as $\frac{5}{747}w$, we get $w.(X.X.w^{-1}) \sim I$
as expected. So, for pure Liouville supergravity model we indeed find that the
operator $w$ is invertible. It is therefore, clear that a general physical
operator for this model can be written as $w^{-n}x^iy^j$, with $0\leq i\leq
p-1$, $0\leq j\leq q-1$ and $n\in Z$. As a final comment, we like to mention
that $w$ has ghost number $-2$ and so, under the ring multiplication the ghost
number of the powers of $w$ will increase by even number, but that will be
inconsistent with the Lian-Zuckerman theorem (15). So, in order to be
consistent one again has to use the picture changing isomorphism [21].

To conclude, we have argued kinematically that the complete cohomology ring of
the Lian-Zuckerman type operators for ${\hat c}_M < 1$ NSR string theory is
generated by two ground ring generators alongwith an additional generator.
By making use of the Lian-Zuckerman theorem in supersymmetric case we
have explicitly constructed the ground ring generators which have ghost
numbers $-\frac{1}{2}$ and belong to R-sector. We have also noted that, there
is a subring of this ground ring generated by three elements in NS-sector
with an equivalence relation amomg them. The additional generator has ghost
number one less than that of the ground ring generators and is invertible.
So, it generates all the higher ghost number operators. This generator belongs
to either in NS-sector or in R-sector depending on whether we consider (even,
even) or (odd, odd) series coupled to $2d$ supergravity. We have noted that in
general, the construction of this operator is quite involved since one has to
deal with higher level null vectors. So, we chose (2, 4) model coupled to
$2d$ supergravity for illustration and constructed this operator. We have
shown by using picture changing isomorphism that it is indeed invertible and
therefore, generates all the higher ghost number operators. It would be nice
to construct a general proof for the invertibility of the additional generator
as argued for the bosonic case in ref.[8]. We have not been able to provide
any because there is no explicit supersymmetric generalization of Felder's
construction. It would also be interesting to explore the symmetry algebra
that underlies ${\hat c}_M < 1$ NSR string theory through the construction of
symmetry charges both for the open as well as for the closed string
cases.
\parbigskip
\parbigskipn
{\bf Acknowledgements:}
\parmedskip
One of us (S.R.) would like to thank A. V. Ramallo for discussions and for
pointing out some references regarding supersymmetric minimal models. The
work of S.R. is supported in part by the Spanish Ministry of Education. S.P.
would like to acknowledge ICTP, Trieste for supporting a visit to the Centre
and the hospitality during which this work was initiated.
\parbigskip
\parbigskipn
\noindent{\bf REFERENCES:}
\parmedskip
\begin{enumerate}
\item D. J. Gross, I. R. Klebanov and M. J. Newman, \np 350 (1991) 621.
\item D. J. Gross and I. R. Klebanov, \np 359 (1991) 3.
\item B. H. Lian and G. J. Zuckerman, \pl 254 (1991) 417; \pl 266 (1991) 21;
\cmp  145 (1992) 561.
\item A. M. Polyakov, \mpl 6 (1991) 635.
\item I. R. Klebanov and A. M. Polyakov, \mpl 6 (1991) 3273.
\item E. Witten, \np 373 (1992) 187.
\item E. Witten and B. Zwiebach, \np 377 (1992) 55.
\item H. Kanno and M. H. Sarmadi, \ijmp 9 (1994) 39.
\item See also S. Panda and S. Roy, \pl 306 (1993) 252.
\item D. Kutasov, E. Martinec and N. Seiberg, \pl 276 (1992) 437.
\item P. Bouwknegt, J. McCarthy and K. Pilch, preprint CERN-TH-6279/91;
\np 377 (1992) 541.
\item K. Itoh and N. Ohta, \np 377 (1992) 113.
\item M. A. Bershadsky, V. G. Knizhnik and M. G. Teitelman, \pl 151 (1985) 31.
\item D. Friedan, Z. Qiu and S. Shenker, \pl 151 (1985) 37.
\item H. Eichenherr, \pl 151 (1985) 26.
\item M. Kato and S. Matsuda, Adv. Stud. Pure Math. 16 (1988) 205.
\item G. Mussardo, G. Sotkov and M. Stanishkov, \pl 195 (1987) 397; \np
305 (1988) 69.
\item G. Felder, \np 317 (1989) 215; erratum, \np 324 (1989) 548.
\item D. Friedan, E. Martinec and S. Shenker, \np 271 (1986) 93.
\item M. Ito, T. Morozumi, S. Nojiri and S. Uehara, Prog. Theor. Phys. 75
(1986) 93.
\item G. Horowitz, R. Myers and S. Martin, \pl 218 (1989) 309.
\end{enumerate}
\vfil

\eject
\end{document}